%% file: main.tex
\documentclass[acmlarge]{acmart}

\setcopyright{acmcopyright}
\copyrightyear{2021}
\acmYear{2021}
\acmDOI{10.1145/1122445.1122456}

\usepackage{makecell}
\usepackage{graphicx}
\usepackage{wrapfig}
\usepackage{subcaption}
\usepackage{listings}
\usepackage{xcolor}

\definecolor{codegreen}{rgb}{0,0.6,0}
\definecolor{codegray}{rgb}{0.5,0.5,0.5}
\definecolor{codepurple}{rgb}{0.58,0,0.82}
\definecolor{backcolour}{rgb}{0.95,0.95,0.92}

\lstdefinestyle{mystyle}{
    backgroundcolor=\color{backcolour},   
    commentstyle=\color{codegreen},
    keywordstyle=\color{magenta},
    numberstyle=\tiny\color{codegray},
    stringstyle=\color{codepurple},
    basicstyle=\ttfamily\footnotesize,
    breakatwhitespace=false,         
    breaklines=true,                 
    captionpos=b,                    
    keepspaces=true,                 
    numbers=left,                    
    numbersep=5pt,                  
    showspaces=false,                
    showstringspaces=false,
    showtabs=false,                  
    tabsize=2
}

\usepackage{algorithm}
\usepackage{algpseudocode}

\input{macros.tex}
\title{Automatic Generation of Interpolants for Lattice Samplings: Part II ---  Implementation and Code Generation}
\author{Joshua Horacsek}
\email{joshua.horacsek@ucalgary.ca}
\affiliation{%
  \institution{University of Calgary}
  \streetaddress{2500 University Dr. NW}
  \city{Calgary}
  \state{Alberta}
  \postcode{T2N 1N4}
}
\author{Usman Alim}
\email{ualim@ucalgary.ca}
\affiliation{%
  \institution{University of Calgary}
  \streetaddress{2500 University Dr. NW}
  \city{Calgary}
  \state{Alberta}
  \postcode{T2N 1N4}
}

\begin{abstract}
In the prequel to this paper, we presented a systematic framework for processing spline spaces. In this paper, we take the results of that framework and provide a code generation pipeline that automatically generates efficient implementations of spline spaces. We decompose the final algorithm from Part I and translate the resulting components into LLVM-IR (a low level language that can be compiled to various targets/architectures). Our design provides a handful of parameters for a practitioner to tune --- this is one of the avenues that provides us with the flexibility to target many different computational architectures and tune performance on those architectures. We also provide an evaluation of the effect of the different parameters on performance.
\end{abstract}



\begin{document}



\maketitle


\input{sections/introduction.tex}
\input{sections/relatedwork.tex}

\input{sections/background.tex}

\input{sections/methodology.tex}
\input{sections/results.tex}
\input{sections/conclusion.tex}


\bibliographystyle{apalike}
\bibliography{main}

\end{document}

%% file: sections/introduction.tex
\section{Introduction}
In Part I, we introduced a framework of analysis that allows one to process a piece-wise polynomial interpolant (defined over a convex simplicial complex) into a list of related ``sub-regions'' with desirable properties for implementation within a machine~\cite{part1}. The implementation details, however were left up to interpretation; this work elaborates upon the generation and evaluation of code from the form given in the prequel. In reality, there are parameters to tune depending on the details of the machine that one wishes to target. For example, code with branching patterns are potentially detrimental to performance on GPU architectures, whereas CPU architectures have been engineered so as to minimize much of the overhead of branching. The main contribution of this paper is a framework for generating code from the abstract description provided in the prequel, and  provide an exploration of the parameter space. While the focus in the prequel was on generality (i.e. interesting test cases), the discussion in this paper is more tuned towards performance.

The analysis in the prequel provides one with a means to explore many different combinations of basis functions and lattices --- translating such a description into a concise and performant implementation is non-trivial. A na\"{i}ve approach is to simply take an interpolant $\varphi$ and evaluate it within the convolution sum: 
\begin{equation}\label{eq:conv_sum}
    \sum_{\mathbf{n} \in L\mathbb{Z}^s} c_{\mathbf{n}}\varphi(\mathbf{x}-\mathbf{n}).
\end{equation}
While this is straightforward, it is plagued by performance issues. For a spline with $n$ points in its support, this leads to $n$ evaluations of the interpolant $\varphi$ for a given reconstruction, however, there is often a good deal of symmetry in higher dimensional splines, thus most of these evaluations contain some amount of repeated work (under symmetry). Moreover, many interesting splines used in scientific visualization are non tensor product splines and thus have many separate cases one needs to consider for different areas of the spline ---  the area a point resides within must be determined at each of the $n$ evaluations of the interpolant. It is more efficient to unroll the convolution sum so as to reduce the amount of repeated work and reduce branching behaviour. 

To accomplish this, while still balancing generality and performance, we propose a code generation framework that takes the abstract representation from Part I, and generates a low level representation that is both close to machine code and relatively platform independent. The main tool we use to accomplish this is the Low Level Virtual Machine (LLVM) which is a framework for code analysis and generation. LLVM is the foundation of many contemporary compilers --- at its core it is a library for expressing low level code; code that is independent of a target architecture. To this end, this allows one to divorce optimisation from high level language design, and allows language designers to focus simply on generating good low level code. Optimization passes are performed solely on LLVM's low level instruction representation (LLVM-IR).

As in Part I, our goal is to make the benefits of non-Cartesian computing more tangible. This work finalizes the ideas of Part I. To summarize, the contributions of this work are as follows:
\begin{itemize}
    \item We provide a unified framework that generates code for fast interpolants from non-separable interpolants on common lattices in scientific computing and visualization. Our framework is generalizeable to $s$-dimensions.
    \item  We detail the parameters of said code generation, elaborate upon how they affect performance for certain types of compute architectures, we also provide a reference implementation which is available on Github~\cite{fastsplinegit} which contains the code to perform the analysis detailed in Part I, the code genertion framework presented in this work, as well as example code that demonstrates the use of the framework and builds upon examples in this work.
    \item We explore the effect of parameter choice on generated code for different compute architectures. Additionally, we explore the performance of the Voronoi splines in 3D, which are arguably the most splines that are closest to the tensor product spline on the Cartesian lattice (in both support size and order.) Prior to this work, they have not received an efficient implementation and have been seen as impractical interpolants.
\end{itemize}
The remainder of this paper is organized as follows.
In Section~\ref{sec:related_work} we touch upon some related work --- this ranges from splines used in 1,2 and 3 dimensional signal processing, to domain specific compilers. We then describe the neccesary background in Section~\ref{sec:background}, referring to the first part for details on shift invariant spaces; it is assumed that the reader is familiar with the first part of this work. In Section~\ref{sec:method}, we detail the mechanisms behind our code generation framework, breaking the system down into components or ``blocks'' and showing how to compose them so as to implement the algorithm in Part I. Finally, in Section~\ref{sec:results} we compare and contrast the effect on performance of different parameters for different compute architectures. 

%% file: sections/relatedwork.tex
\section{Related Work} \label{sec:related_work}
The motivating factor for this work is the plethora of other publications investigating non tensor product splines in scientific visualization. However, there are also many works that investigate novel interpolants in lower dimensions. For uni-variate splines, the optimal class of splines in the maximal order minimal support (MOMS) splines~\cite{blu2001moms}, although there are some notable, yet non-polynomial, splines such as the CINAPACT splines that boast nice properties such as being infinitely differentiable~\cite{cinapact}. Authors tend to neglect implementation details for uni-variate splines since it is typically straightforward --- simply fix the ``region of evaluation'' as a line segment, then distribute the polynomials from the convolution sum; polynomial evaluation is performed (optimally) with Horner's algorithm~\cite{horner1}. In 2D, there has been some work around hexagonal interpolants, however, the provided implementations exist as MATLAB code~\cite{condat2006three}.

Implementing a shift invariant reconstruction space becomes more tedious as the dimension of the interpolant increases. If one wishes to implement tensor product schemes, it is relatively straightforward to extend univariate implementations by simply reducing the problem along each dimension. If one is considering non tensor product splines, and/or non-Cartesian lattices, the problem becomes more difficult. The reason for this is that non tensor product splines are non-separable --- one can no longer simply apply univariate results along each axis. Typically, the problem requires some amount of geometric analysis to take advantage of the symmetry of a spline. As such, when an implementation appears in the literature, it may include a detailed implementation, but more often it is followed by a publication describing an efficient implementation~\cite{kimeval,finkbeiner2010efficient,csebfalvi2013cosine}. However, these implementations are, again, often specific to a single language and platform. We aim to fill that gap in this paper, and the main tool we use is LLVM. Specifically, we take an abstract representation of a shift invariant reconstruction space and translate it into a low level representation. 

There are many benefits in translating to LLVM-IR. First of all, it allows us to represent a reconstruction space with low level code that is relatively agnostic to the underlying machine architecture. This strong abstraction between high-level and low-level representations better facilitates further processing of the original abstract representation. Languages such as Julia keep their abstract representation wholly distinct from the underlying machine~\cite{bezanson2017julia}; one may build additional processing steps on top of this representation, such as auto-differentiation, then pass the resulting representation through the same code generation mechanism used for the original code. The second main benefit we reap is the removal of the overhead that comes from a general purpose implementation of a shift invariant reconstruction space. While there is no other work that specifically attempts to take a shift invariant space and generate code, there are works within numerical computing that use LLVM as a tool to reduce overhead in a similar way as described below.

TensorFlow's XLA (Accelerated Linear Algebra) library is a domain specific compiler\footnote{From this perspective, one may consider our work a domain specific compiler whose input is a spline reconstruction space.} that translates components of a computational graph into LLVM-IR, which is in turn compiled to a target architecture~\cite{xla2017xla}. The main benefit XLA provides is that one may fuse together operations in a computational graph so as to keep more operations in registers during computation, thereby reducing overhead by using less memory bandwidth. While the improvement is modest; roughly 1.14$\times$ in their benchmarks, the improvement is consistent, especially when one considers that many applications that use TensorFlow are long running applications. There are also libraries, such as Nimble~\cite{shen2020nimble}, that compile deep learning models down to lower level code to both reduce the overhead of inference, and remove dependencies on large deep learning libraries.

From the perspective of numerical computation, domain specific compilation is a relatively new, but powerful, technique. Combined with an appropriate analysis framework, it has been used to accelerate matrix operations --- this is particularly useful when the matrices in question have very specific structure~\cite{fabregat2012domain}. Another example of numerical computation domain specific compiler (and language) is SDSLc, a language for generating fast stencil computation code~\cite{rawat2015sdslc} --- curiously, this work does not acknowledge or make use of LLVM.

%% file: sections/background.tex
\section{Background} \label{sec:background}
\begin{figure}[t]
  \centering
    \begin{subfigure}{0.25\textwidth}
      \includegraphics[scale=1.6]{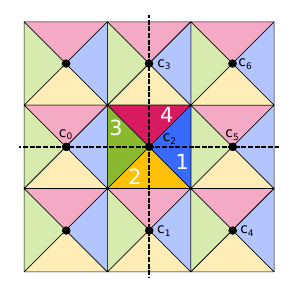}
    \end{subfigure}
    \begin{subfigure}{0.74\textwidth}
      \centering
        \begin{tabular}{cccc}
        Sub-region & Polynomial           & Transform  \\ \hline \hline
        1          & \makecell{
        $\frac{1}{4} \, {\left(2 \, c_{0} - c_{1} - 2 \, c_{2} - c_{3} + c_{4} + c_{6}\right)} x_{0}^{2} + $ \\  
        $ \frac{1}{4} \, {\left(c_{1} - 2 \, c_{2} + c_{3} + c_{4} - 2 \, c_{5} + c_{6}\right)} x_{1}^{2} +$ \\ $ \frac{1}{2} \, {\left({\left(c_{1} - c_{3} - c_{4} + c_{6}\right)} x_{0} - c_{1} + c_{3}\right)} x_{1} + $ \\ $ \frac{1}{8} \, c_{0} + \frac{1}{8} \, c_{1} + \frac{1}{2} \, c_{2} + \frac{1}{8} \, c_{3} + \frac{1}{8} \, c_{5} -$ \\ $ \frac{1}{2} \, {\left(c_{0} - c_{5}\right)} x_{0}$} & None              \\
        2          & Reference to sub-region 1             &  $\circlearrowright 90^\circ$                  \\
        3          & Reference to sub-region 1              &         $\circlearrowright 180^\circ$             \\
        4          & Reference to sub-region 1              &   $\circlearrowleft 90^\circ$              \\ \hline
        \end{tabular} 
    \end{subfigure}
    \caption{The image on the left continues the example from Part I, showing the sub-region decomposition for the ZP-element~\cite{part1}. The square region in the center is the region of evaluation; it is subdivided into the sub-regions of evaluation, labelled 1-4. The coefficients $c_0$ to $c_6$ denote the lattice sites that contribute to a reconstruction in sub-region 1. The table on the right details the relationship between sub-regions --- the transformation is generally affine, however in this case we do not need to make use of any shift. Sub-regions with no transform are {\rm reference} sub-regions, in this example there is only one. However, in general a spline may emit more than one reference sub-region (e.g. the cubic spline on the FCC lattice~\cite{kim2008box}). }
    \label{fig:ex1}
\end{figure}

Our focus in Part I was to present the theoretical foundation needed to transform the defining objects of a shift invariant space (a lattice and a basis function) into a form that can easily be translated into an implementation --- it is assumed that the reader has a good notion of the content and background presented in that work~\cite{part1}. The translation into LLVM is the focus of this paper. In the prequel, our analyses provided us with a shift invariant structure over a lattice; in specific, we concluded with a shift invariant polyhedral region of evaluation, and a dissection of the region of evaluation into sub-regions of evaluation. One only needs to consider a single space-tiling region of evaluation since, by definition of a shift invariant space, every other region of evaluation is related by a shift. Each sub-region contains a polynomial and a set of lattice sites. The data at these lattice sites are components of the polynomial belonging to each sub-region and the polynomial dictates how to reconstruct a value in a given sub-region. Figure~\ref{fig:ex1} visualizes all these components. 

Very often in practice, the splines we use on a given lattice are symmetric, at least in dimensions 1 to 3 --- this helps encourage reconstruction spaces to capture frequency content more evenly in all directions. This is because the underlying shift invariant space inherits, partially, the symmetry of the basis function --- in reality, the symmetry of the underlying space is a combination of the symmetry of the lattice and the basis function. This symmetry is a tool that allows us to write different sub-regions in terms of each other --- relating them by rigid body transformations, thereby allowing us to reduce the amount of polynomials stored for each sub-region. For most splines used in visualization, we only require one distinct polynomial, this polynomial is then used to reconstruct values within each sub-region (after an appropriate change of variables). As such, we also store this auxiliary information. That is, if there exists a rigid body transformation that takes one region into another, we only store that transformation.

The above representation is generic and independent from the machine on which we will implement said scheme. This is advantageous because it allows for further simple processing of the spline space. One may easily take derivatives of the polynomials within sub-regions to obtain gradient estimates (note that the symmetry analyses still apply even for derivative computation). This can be taken further to compute the Hessian, provided the spline is smooth enough. One may also transform these polynomials into forms that are more convenient for evaluation, i.e. Horner form, or Berstein B\'{e}zier (BB) form. 

A na\"ive approach, provided this representation, would be to represent each polynomial into a specific form --- BB-form for example --- then implement an algorithm to perform that style of polynomial evaluation. For polynomials in BB form, the algorithm would be De CastleJau's algorithm. However, the algorithm we choose may be inappropriate for a given machine. For example De CastleJau's algorithm is recursive, which is difficult to implement on GPU architectures --- this can, in part, be alleviated by unrolling the recursion. Another downside of simply using  De CastleJau's algorithm is temporary space usage. Temporary space usage is bound by the number of terms in the BB form (i.e. the degree of the polynomial). On modern CPU architectures, this space is allocated on the stack, which resides in main memory, but is likely aliased to cache for fast access. On the GPU, temporary storage space often means increasing register allocation for a program, which can drastically reduce the amount of parallelism available for the program. Alternatively, higher temporary storage needs may cause registers to ``spill'' into global memory, requiring global memory accesses which are very slow. Moreover, polynomial evaluation is only one part of the reconstruction scheme, there is also the preamble of determining which region and sub-region of evaluation a point resides in prior to reconstruction. This must be accounted for --- there is overhead in looking up the properties of a given reconstruction space and delineating what is to be done in certain cases. 
 
The approach we take is to generate implementations rather than provide a single implementation that takes a lattice and a basis function as parameters. This allows us to reduce some amount of overhead by ``baking'' information into the generated code. For example, we are able to remove the (small) cost associated with looking up member variables, hard-code any logic that may need to be applied for a specific basis function or lattice, and unroll loops as necessary. This is similar to template meta-programming in C++ which, in part, exists to take advantage of such optimizations. Code generation also allows us to replace certain components of our pipeline with other implementations. For example, we may pull out the Horner polynomial evaluation and replace it with De Castlejau's algorithm. We avoid doing so in this work, since Horner schemes have low temporary storage requirements and are easily unrollable. There are also many options for decomposing and pipelining a Horner scheme, which we also extensively explore --- the same treatment for De Casteljau's algorithm is outside the scope of this paper. 

\subsection{Machine Architecture}
One simple approach to code generation is to choose a popular language, such as C, as a target language, then create ``templates'' for specific parts of the reconstruction function. For example, one component may be to determine the region of evaluation, whereas another may be to evaluate a given polynomial. In our experience, this becomes messy quite quickly, the amount of cases one must write and test is fairly large. Moreover, the template code becomes unreadable quickly. This approach also lacks elegance; we would start with a low level description of the spline space, then generate high level code, which must be processed by a compiler to produce low level code. A more elegant approach is to directly generate a low level representation. Thus, we emit code to LLVM intermediate representation (LLVM-IR) directly; this has the benefit of being low level but still platform and machine agnostic.

\begin{figure}[t!]
\centering
\includegraphics[scale=0.7]{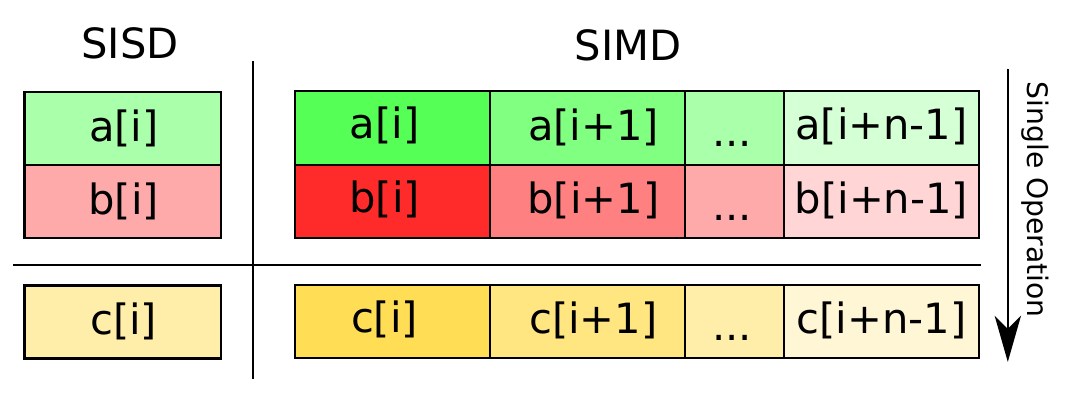}
\caption{An example of the difference between SISD and SIMD architecture. Here, $n$ is the vector length, and as a single operation is applied to the operands $a$ and $b$, this operation is applied in parallel to all elements of $a$ and $b$.}
\label{fig:machineex}
\end{figure}

One must, however, take into account {\em some} differences in machine architecture when generating code. According to Flynn's taxonomy~\cite{flynn1972some}, there are four main classes to computer architecture. The two of interest to us are {\em single instruction single data} (SISD) machines in which each operation operates on at most one element of data, and {\em single instruction multiple data} machines (SIMD) in which an operation may operate over multiple elements of data. The other two main distinctions: {\em multiple instruction multiple data} (MIMD) and {\em multiple instruction single data} are not as relevant --- from the perspective of a compute thread, instructions are either SISD or SIMD. Figure~\ref{fig:machineex} show the differences between the two.
Most modern CPUs are super-scalar machines, that is, they perform instruction level parallelism --- many operations may be in flight at any given time. For example, a memory fetch may appear early in an instruction stream, but may only be required near the end of the stream; the CPU may commit this to the memory controller and move on to the next instruction, only stalling once the required memory access is actually needed and has not finished yet. This is in contrast to traditional scalar architectures, in which instructions are strictly executed in order and have predictable execution times. From the perspective of a thread, most contemporary CPUs are SIMD machines, whose SIMD instructions are vectorized SISD instructions.

\subsection{GPU Architecture}
GPUs have historically been fixed function multi-processing units dedicated to triangle rasterization --- that is, a single GPU contained a large amount of cores dedicated to transforming and projecting triangles onto a planar display. This allowed for a high level of paralellism for the task of rasterization. However, as the graphics pipeline became more programmable through the use of different types of shaders, GPUs eventually became general purpose computing devices. Modern GPU architectures are more akin to general purpose SIMD machines with wide instruction lanes --- a single instruction on a GPU will operate on 32 to 64 elements of data, depending on the manufacturer. While this is true at the hardware level, the programming model presented by GPU manufacturers is that each lane of execution corresponds to a lightweight ``thread''. From the perspective of each thread, the instruction stream is largely SISD. 

Again, historically practitioners have taken advantage of the parallelism offered by GPUs through OpenGL shaders. Shaders in OpenGL are small programs that execute on the GPU and operate element wise over data: pixels, vertices and/or triangles. However, this quickly became insufficient for general purpose computing on the GPU (GPGPU). This was for various reasons, but one pressing reason was the lack of fine grained control over hardware resources. NVIDIA first offered a solution to this, CUDA, which provided better control over (and access to) the underlying hardware. Programs in CUDA are known as CUDA kernels and execute element-wise over data. CUDA code runs on CUDA devices, these devices contain their own address space and are connected to a host via a PCI-e bus. Data and programs must be uploaded across the PCI-e bus prior to execution. Once a kernel is completed, the results of the computation must be again copied off the device into host memory.

The basic unit of execution in CUDA is a thread; threads are collected in groups of 32 called warps. Warps execute programs in lockstep and share the resources of a streaming multiprocessor (SM) --- that is, an SM has a set of registers that are allocated to each thread in a warp; based on the problem one wishes to solve, more or less registers may be required. Threads in a warp may communicate with each other at the register level, and may use fast local shared memory to communicate with other threads within the same block. The amount of shared memory and registers allocated per thread may be so that it is not possible to schedule a warp on each core of a SM at any given moment. The ratio of active warps to total number of possible warps active is known as occupancy. Generally, one wishes to maximize occupancy to make the best use of resources on a given SM, but it is not always possible to do so.

Grids are collections of blocks and are the most coarse grain level of problem decomposition. For example, to render an image, we may break it up into chunks of 8x8 pixels, or a total of 32 pixels. This would correspond to a block size of 64 (block sizes must be multiples of the warp size, i.e. 32). There is no guaranteed order in which blocks complete --- for the example above completion order is irrelevant, one simply needs to ensure that all blocks are completed before the image is displayed. If synchronization is needed between blocks, then one must synchronize by waiting for the current kernel to finish, then launching a new kernel\footnote{Although block level synchronization is supported through cooperative groups in CUDA 9 and above.}. 

Since GPU cores are SIMD machines at the hardware level, one potential difficulty with the lightweight thread model of CUDA is branching --- what are threads (i.e. SIMD lanes) to do when a branch occurs? Provided all threads take the same path for a branch, then there is no issue. If threads within a warp take different paths at a branch, then those threads are disabled and revisited later. This is known as branch divergence. This is something one generally wishes to avoid, at least for long runs of instructions. This can be dealt with by algorithm design or branch predication.

From the perspective of an application running on the GPU, to compute a value from an interpolation scheme as fast as possible, one may be tempted to distribute computation along each lane of a warp in terms of SIMD threads. However, this approach makes little sense, as applications are likely demanding reconstructions on a per thread basis. Thus it makes more sense to create optimized SISD instruction streams along every lane. However, taking the same observation into account for code on the CPU, again each thread is likely going to correspond to one element of the problem, but each thread also has access to SIMD vector instructions. Additionally, one must take care to use the appropriate memory fetch --- on a GPU texture fetches are spatially cached, and one must take care to organize memory reads to take advantage of this.

Of course NVIDIA is not the only GPU vendor, AMD also produces comparable GPUs, but the eco-system for developing GPGPU applications on AMD has been limited to OpenCL until recently with HIP. The concepts in OpenCL are very similar to those in CUDA --- the concept of a warp translates to that of a wavefront (which is a collection of 64 threads); block translates to work group; and grid translates to overall problem size.

Another subtlety of GPU compute applications is that bandwidth is a more prominent serious concern. Since compute throughput is high, GPUs must be able to read and write data to memory at comparable rates. To facilitate this, the best performance is achieved when memory access patterns are simple. CUDA provides a programmer with various different levels of memory. The first is a register; registers which are accessible to a thread. The next level is constant memory and shared memory, then global memory. Each is an order of magnitude slower than the last.

\subsection{LLVM}
Shifting a performant implementation of any algorithm to a different architecture can be a daunting task. In that respect LLVM is a powerful tool. One of the core components of LLVM that helps facilitate this is LLVM-IR, a low level assembly-like language targeting a fictional virtual machine. This representation is in single static assignment (SSA) form. That is, values are assigned to output variables, then those variables cannot be modified, and may only be used as inputs in subsequent operations. While LLVM-IR can be interpreted and therefore explicitly executed, doing so is only advisable when the LLVM tool-chain does not support code generation on a specific architecture; the intention of LLVM-IR is to be translated into machine code of a given architecture. This can either be linked into an executeable at compile time, or linked into a running executable on-the-fly (i.e. just in time or JIT compilation).
\begin{figure}
    \centering
    \includegraphics[scale=0.7]{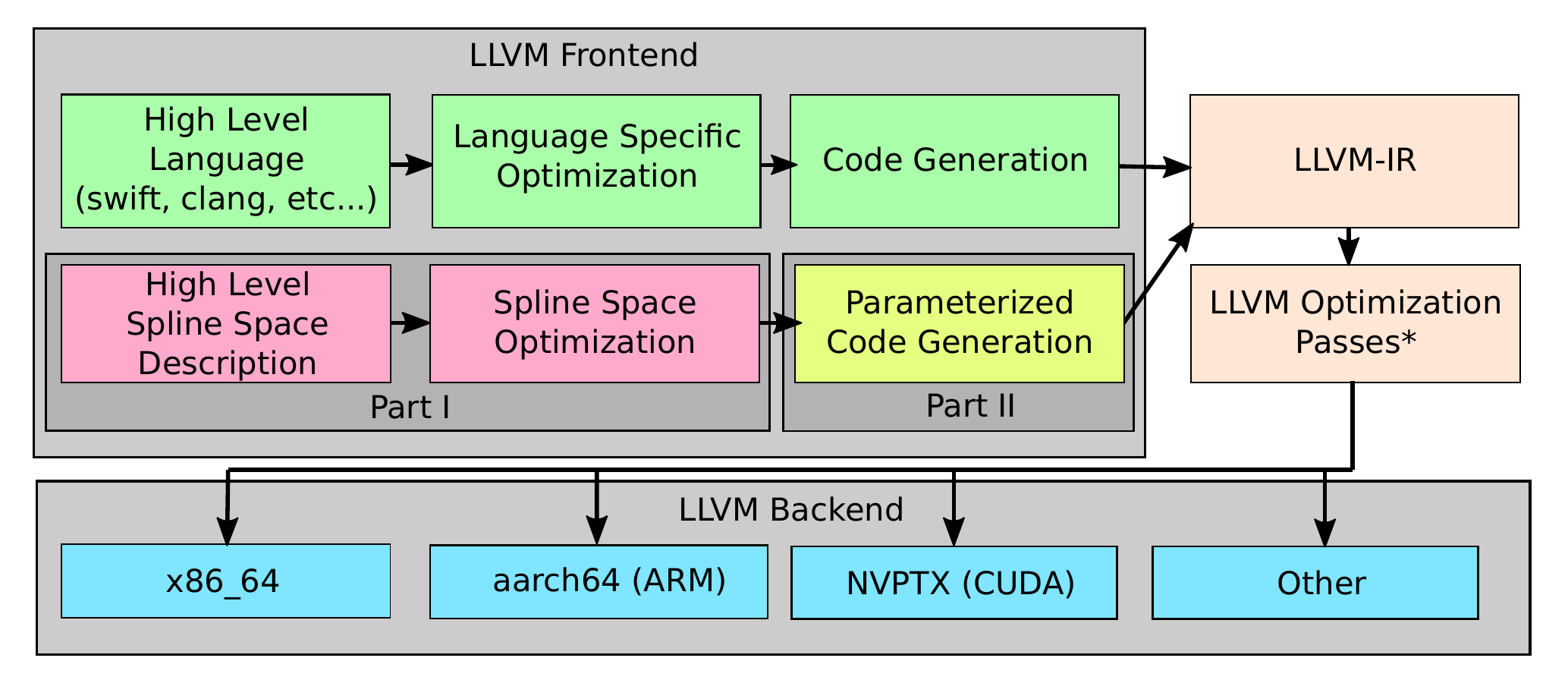}
    \caption{A high level overview of the LLVM ecosystem, including our analysis and code generation contributions. The * in the LLVM optimization passes denotes that they are optional.}
    \label{fig:llvmhl}
\end{figure}

In terms of code generation, it is the responsibility of the {\em backend} translator to take LLVM-IR and generate machine code. For the most part, the job of a backend translator is to assign registers and stack space to operands and results, as well as translating LLVM-IR instructions to the target machine's architecture. Prior to final compilation, LLVM also provides many optimization passes --- these passes transform LLVM-IR into optimized LLVM-IR. The decision to implement optimization passes at the level of LLVM-IR reduces the complexity of compilers producing LLVM-IR (also known as {\em frontends} for LLVM; frontends need not worry too much about optimization. Figure~\ref{fig:llvmhl} shows how the frontend, optimization passes, and backend relate, as well as where our works fit into the LLVM ecosystem.

LLVM is a natural complement to the CUDA ecosystem. CUDA similarly provides a low-level machine abstraction, the parallel thread executable (PTX) environment. PTX code reads very much like LLVM-IR however with more low-level constructs that expose certain features of the hardware to a developer (thread communication primitives, for example). PTX and LLVM are similar in their goals --- PTX exists as a means of representing shader/compute code in a low-level format, yet still affords some platform agnosticism. 

Again, PTX code is written from the perspective of a single thread, so translating LLVM-IR to PTX is relatively straight-forward. Final register allocation, grid size, and block size will be determined by the CUDA run-time, and is independent of the LLVM-IR/PTX representation of the code. In the next section, we detail how we use LLVM to translate each component of a reconstruction space into LLVM-IR, and the different variations on how each component is parameterized.

%% file: sections/methodology.tex
\section{Methodology} \label{sec:method}
We start by breaking down the components of the algorithm proposed in Part I into separate distinct chunks, then elaborate upon how code is generated for each of these chunks. We reiterate the algorithm here with minor modifications; specifically we make the memory fetches explicit in Algorithm~\ref{alg:eval}. We first decompose Algorithm~\ref{alg:eval} into different parts: the preamble, the summation loop and polynomial evaluation.

We present these as different generation ``blocks''. Blocks take in LLVM-IR variables, and output a set of LLVM-IR variables. Some input variables may correspond to input (i.e. the point at which we are evaluating the spline) whereas other may correspond to internal use elements (lookup tables, or the output of other blocks). Bridging the gap between input and output are LLVM-IR instructions which take input, and provide one output. We proceed with the running example from Part I, which has already been reintroduced in Section~\ref{sec:background}.

\begin{algorithm}
    \caption{Branch free evaluation at a point.}
    \label{alg:eval}
    \begin{algorithmic}[1] 
        \Procedure{Eval}{$\mathbf{x}$}
            \State $f \gets 0$
            \For{$\mathbf{l} \in \{\mathbf{l}_0,\mathbf{l}_1,\cdots \mathbf{l}_{M-1}\} $} 
            
            \State $\mathbf{k} \gets \rho(\mathbf{x})$  \Comment{Determine the shift to ROE}
            \State $\mathbf{x} \gets \mathbf{x} - \mathbf{k}$   \Comment{Shift ROE}
            \State $q \gets 0$
    
            \For{$i \in \{0,1,\cdots Q-1\} $} \Comment{Determine BSP index}
               \State $q \gets (\mathbf{x}\cdot\mathbf{p}_i - d_i < 0) \ ? \ q : q \mathbin{|} 2^i$
            \EndFor
            \State $SubRegionIndex \gets \sigma(q \% p)$\Comment{Map BSP index into sub-regions}
            
            \State $g \gets 0$
            \State $T^\prime \gets T[SubRegionIndex]$
            \State $\mathbf{t}^\prime \gets -T^\prime\mathbf{t}[SubRegionIndex]$
            \State $\pi^\prime \gets \pi[SubRegionIndex]$
            \For{$j \in \{0,1\cdots, n-1\}$}
               \State $c_j \gets MemoryLookup(\pi^\prime[LatticeSite[j]])$ 
            \EndFor 
            \For{$i \in \{0,1,\cdots, K-1\} $} 
                \State $v \gets {\psi_i^{\pi^\prime}}(T^\prime\mathbf{x} - \mathbf{t},  \mathbf{k}) $
                \State $g \gets g + PsiIndex[SubRegionIndex] == i \ ? \  v \  : \  0$
            \EndFor
            
            \State $f \gets f + g$\Comment{Add the contribution for this coset}
            \EndFor
            \State \textbf{return} $f$ 
        \EndProcedure
    \end{algorithmic}
\end{algorithm}

\subsection{Preamble}
The generated code begins with a function declaration and generation of the lookup tables needed for computation --- specifically, the tables precomputed in Part I (explicitly $T, \mathbf{t}, \pi$ and $PsiIndex$ in Algorithm~\ref{alg:eval}) are populated if necessary; some splines have only one element in each of these tables, so tables are not generated in those cases and that single element will be inlined when needed. The function definition takes a spatial location (i.e. $s$ floating or double scalar values) and a memory lookup primitive. The lookup primitive can be a pointer to a function, a sequence of memory arrays (each corresponding to a Cartesian coset of the lattice) or a texture object (for GPU code). If the lookup primitive is a function pointer, this function is called on every lattice site lookup. If an array is specified, memory fetches are generated at every lattice site fetch. If the lookup primitive is a texture fetch, then code generation module will generate texture fetches when the data within lattice sites are needed. These can be specified as either linear or nearest neighbor fetches. Listing~\ref{lst:preamble} shows an example of the generated LLVM-IR code containing lookup tables and the function declaration.
\begin{lstlisting}[language=LLVM, label={lst:preamble}, caption=Example of function  and lookup table definition. Some splines may have more or less lookup tables; some have none. In practice `texture\_t' is presented as an i64 in order to remain transparent to LLVM.]
; Different indexes are declared here, addrspace(4) tells the backend to place these tables in constant memory
@"bsp_index" = addrspace(4) constant [8 x i8] ;; constant data omitted
@"xform_lookup" = addrspace(4) constant ; constant data omitted
; For functions passed a pointer, the function signature is: 
define double @reconstruct(double %x_0, double %x_1, double (i32, i32)* %lookup ) {
; For functions passed a texture, the function signature is: 
define double @reconstruct(double %x_0, double %x_1, texture_t %lookup ) {
entry:
  ; ... composition of blocks ...
}
\end{lstlisting}

\subsubsection{Coset Loop}
If the spline space has a coset decomposition, we generate the header for a loop over the cosets. We first generate a label and index, then generate the memory look-ups for the coset offsets. We then increment the index for the next iteration. After the subsequent blocks, a conditional branch brings execution flow back to the label we defined above.

\subsubsection{Region of evaluation}
\begin{figure}[h!]
    \begin{subfigure}{0.2\textwidth}
      \includegraphics[scale=1.3]{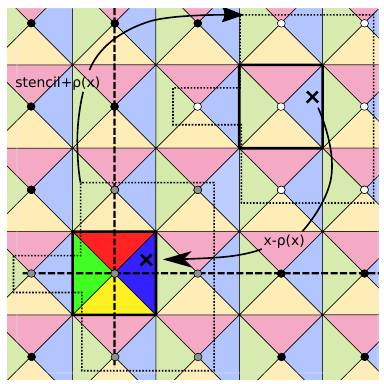}
    \end{subfigure}
    \hspace{2.5cm}
    \begin{subfigure}{0.63\textwidth}
\begin{lstlisting}[language=LLVM]
  %tmp0 = fmul float %x_0, 1.0 ; inverse matrix transform. 
  %tmp1 = fmul float %x_1, 0.0 ; these operations will get
  %tmp2 = fmul float %x_0, 0.0 ; optimized away, but are 
  %tmp3 = fmul float %x_1, 1.0 ; kept here for clarity
  %tmp4 = fadd float %tmp0, tmp1 
  %tmp5 = fadd float %tmp2, tmp3 
  ; clamp to center of parallelpiped
  %pre_rho0 = call float @"llvm.round.f32"(%tmp4)
  %pre_rho1 = call float @"llvm.round.f32"(%tmp5)
  %tmp6 = fmul float %pre_rho0, 1.0 ; forward matrix transform.
  %tmp7 = fmul float %pre_rho1, 0.0 
  %tmp8 = fmul float %pre_rho0, 0.0 
  %tmp9 = fmul float %pre_rho1, 1.0 
  %rho0 = fadd float %tmp6, tmp7 ; float, but will be cast 
  %rho1 = fadd float %tmp8, tmp9 ; to integer when needed

\end{lstlisting}
    \end{subfigure}
    \caption{The image on the left exemplifies the utility of $\rho(x)$ --- space is tesselated into regions of evaluations, using $\rho(x)$ we shift it to our reference region of evaluation. In general this is performed by an inverse matrix transform, then either a round or floor operation, and finally a transform back to the original space tiled by the paralellpiped. This is exemplified by the LLVM code on the right. The value of $\rho(x)$ is also used to shift the lookup lattice sites (i.e. the stencil) from the reference region to the correct lattice sites. }
    \label{fig:roe}
\end{figure}

This block determines the lattice site closest to (or within) the region of evaluation. The user may specify the region of evaluation as a paralellpiped region that tiles space according to the given lattice (in which they specify the paralellpiped via a matrix); they may also specify the region of evaluation as the Voronoi cell of the lattice. These are two methods of collecting the reference sub-regions of evaluation, there other possible groupings, but these two are very simple to implement and cover a wide variety of cases. The output of this block is an $s$-tuple of LLVM-IR variables that correspond to a lattice shift to the reference region. The input point is then shifted by the reference point. Figure~\ref{fig:roe} demonstrates how $\rho$ is calculated, and also (geometrically) how it is used.

\subsubsection{Sub-region membership}
\begin{figure}[h!]
    \begin{subfigure}{0.44\textwidth}
      \includegraphics[scale=1.7]{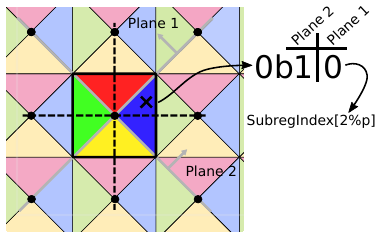}
    \end{subfigure}
    \hspace{0cm}
    \begin{subfigure}{0.55\textwidth}
\begin{lstlisting}[language=LLVM]
; Test against plane 1
%srm0 = fmul float %xt_0, -1.0 
%srm1 = fmul float %xt_1, 1.0 
%srm2 = fadd float %srm1, srm0
%srm3 = fcmp oge float %srm3, 0
%bsp_index0  = select i1 %srm3, i32 1, i32 0
; Test against plane 2
%srm4 = fmul float %xt_0, 1.0 
%srm5 = fmul float %xt_1, 1.0 
%srm6 = fadd float %srm5, srm4
%srm7 = fcmp oge float %srm6, 0
%bsp_index1  = select i1 %srm7, i32 2, i32 0
%bsp_idx = or i32 %bsp_index1, %bsp_index0
; lookup the subregion
%lutidx = urem i32 %bsp_idx, 4
%el_addr = getelementptr [8 x i8], [8 x i8] addrspace(4)* @"bsp_index", i32 0, i32 %bsp_idx
%subreg_idx = load i8, i8 addrspace(4)* %el_addr
\end{lstlisting}
    \end{subfigure}
    \caption{The image on the left shows the planes that decompose the region of evaluation into sub-regions of evaluation. To determine the sub-region, the point of evaluation is shifted to the reference region, then the point of evaluation is tested to determine which side of each plane it lies on. Each plane test corresponds to a bit of an integer, which then is compressed down into a reasonable range before being fed into a lookup table that yields the final sub-region index. The code on the right demonstrates what this procedure looks like in LLVM-IR.}
    \label{fig:bspexample}
\end{figure}

Once the point of evaluation has been shifted to the region of evaluation, the next task is to determine the sub-region the point resides within. The next block takes in the shifted evaluation point from the previous blocks and compares the point of evaluation with the planes that split the region of evaluation into sub-regions. Each comparison consists of a dot product and a subtraction or addition, this is easily vectorized, however the vector reduction functions are still experimental in LLVM, thus we leave this as scalar code, and hope that a given backend will vectorize this code if possible. Each plane comparison corresponds to a bit in an unsigned integer; where the integer width is chosen to accomodate the number of planes, this produces an integer $q$. Figure~\ref{fig:bspexample} shows both how the plane comparisons determine the sub-region index, as well as the generated code. While $q$ could be used to index the sub-region of evaluation, we may have drastically more possible values of $q$ than sub-regions, this would require an entry for ever possible value of $q$ in our lookup tables. Instead, we take $q$ mod $p$ to compress it to a more suitable range, then we pass this through a lookup table that produces the final sub-region of evaluation. 

\subsection{Table Lookups}
Now that the sub-region index is known, we may look up the various transformations needed to translate one sub-region into another. There are a few subtleties to note here. The first is that even though we require only one query to the different lookup tables, it may be advantageous to explicitly re-fetch the elements of these lookup tables whenever they are needed --- this hints at the lifespan of a register and allows the backend to allocate less registers, increasing occupancy, at the cost of a few memory accesses. On the GPU the lookup tables are stored in constant memory and avoid additional slow accesses to main memory; on the CPU these tables will quickly enter the CPU's cache memory system. Since this operation consists only of memory look-ups, we omit an example listing.

\subsection{Reference Sub-region Loop}
The next significant body of code is the loop over all reference sub-regions. This loop is always explicitly unrolled, since the representation of the polynomial may not necessarily be the same per reference sub-region.

\subsection{Polynomial evaluation}
The final step is to evaluate the given polynomial within a sub-region. Since the width of the memory bus on a compute architecture is variable, as is the ``raw compute to memory bandwidth'' ratio of modern architectures, we introduce parameters to tune the performance of the polynomial evaluation on a per-architecture basis. The first parameter we explore is a consequence of dividing the polynomial within a sub-region into smaller chunks; that is, we divide the polynomial into groups of size $m$ based on memory accesses. This allows us to interleave computation and memory accesses  --- that is, as memory read instructions are waiting on results, we are able to compute parts of a polynomial that do not depend on the memory reads still in flight. First, the polynomial is split up into groups that depend only on (at most) $m$ coefficients --- we call $m$ the evaluation {\em group size}. At any given moment, we allow a maximum of $d$ (such that $d \ge m$) memory reads to be sent to the memory controller, computations for the first $m$ coefficients begin once they have come back from the memory controller --- we call $d$ the {\em pipeline depth}. To make this idea more concrete, we demonstrate the principle with the example from the above. We set $m=2$ and $d=4$, then the polynomial in  Figure~\ref{fig:ex1} is decomposed into the following polynomials:
\begin{eqnarray}
p_0(x_0,x_1,c_0,c_1)  & := & \left(\frac{1}{2}c_0 - \frac{1}{4}c_1x_0-\frac{1}{2}c_0+\frac{1}{2}c_1x_1\right)x_0-\frac{1}{2}c_1x_1+\frac{1}{8}c_0 + \frac{1}{8}c_1+\frac{1}{4}c_1x_1x_1 \\
p_1(x_0,x_1,c_2,c_3)  &:= & \left(\frac{1}{2}c_3-\frac{1}{2}c_3x_0-\frac{1}{2}c_2 + \frac{1}{4}c_3x_1\right)x_1-\frac{1}{2}c_2 - \frac{1}{4}c_3x_0x_0+\frac{1}{2}c_2 + \frac{1}{8}c_3 \\
p_2(x_0,x_1,c_4,c_5) &:= & \left(\frac{1}{4}c_4x_0+\frac{1}{2}c_5-\frac{1}{2}c_4x_1\right)x_0+\frac{1}{8}c_5+\frac{1}{4}c_4 - \frac{1}{2}c_5x_1x_1 \\
p_3(x_0,x_1,c_6) &:= & \left(\frac{1}{4}c_6x_0+\frac{1}{2}c_6x_1\right)x_0+\frac{1}{4}c_6x_1x_1
\end{eqnarray}
Each of the above polynomials is in a multivariate Horner form (from a greedy Horner factorization). Summing all of these polynomials gives the polynomial for the reference sub-region in Figure~\ref{fig:ex1}. Keep in mind that each of the $c_i$ correspond to memory lookup at a lattice site, so the decomposition above will depend on the order in which one decides to visit $c_i$ --- for discussion on this and its optimization procedure, see Part I. Algorithm~\ref{alg:pipeline} shows a concrete example of pipelining.
\begin{algorithm}
    \caption{Piplineing example.}
    \label{alg:pipeline}
    \begin{algorithmic}[1] 
        \State $c_0 \gets FETCH( T^{t} [-1,0]^t + [rho0, rho1]^t)$ \Comment{Commit 4 memory reads}
        \State $c_1 \gets FETCH( T^{t} [0,-1]^t +[rho0, rho1]^t)$
        \State $c_2 \gets FETCH( T^{t} [0,0]^t + [rho0, rho1]^t)$
        \State $c_3 \gets FETCH( T^{t} [0,1]^t + [rho0, rho1]^t)$
        
        \State $STALL(c_0)$  \Comment{Wait for 2 fetches to complete}
        \State $STALL(c_1)$ 
        \State $result \gets p_0(xt_0, xt_1, c_0,c_1)$ \Comment{Compute what we can while memory fetches are in flight}
        \State $c_4 \gets FETCH( T^{t} [1,-1]^t + [rho0, rho1]^t)$ \Comment{Commit 2 new reads to the controller before }
        \State $c_5 \gets FETCH( T^{t} [1,0]^t + [rho0, rho1]^t)$
        \State $STALL(c_2)$ 
        \State $STALL(c_3)$ 
        
        \State $result \gets result + p_1(xt_0, xt_1, c_2,c_3)$
        \State $c_6 \gets FETCH( T^{t} [1,1]^t + [rho0, rho1]^t)$
        \State $STALL(c_4)$ 
        \State $STALL(c_5)$ 
        \State $result \gets result + p_2(xt_0, xt_1, c_4,c_5)$
        \State $STALL(c_6)$ 
        \State $result \gets result + p_3(xt_0, xt_1, c_6)$
    \end{algorithmic}
\end{algorithm}

\subsubsection{Polynomial Evaluation with Linear Texture Fetches}
When the linear fetch is used, the order of operations changes compared to the case above. In this case, parts of the polynomial are computed prior to the trilinear memory fetches. Thus the approach is similar to the above, but the order of fetches and computation are reversed. Moreover, it is no longer possible to pipeline polyomial evaluation as above, since the trilinear decomposition requires an evaluation group size of 1.

%% file: sections/results.tex
\section{Experiments and Results} \label{sec:results}
To validate the performance of our method and demonstrate the utility of our parameterization, we measure the average ``speed'' of our generated code for various cases. In this context, we take ``speed'' to mean the average number of point-wise reconstructions one may perform per second. We do this for all valid combinations of pipeline depth, maximum fetch count and, when appropriate, branch behaviour. We also collect and report the variance associated with each experiment when possible.

We performed our tests over three different architectures: x86\_64, ARM, and CUDA. All of our x86\_64 tests were performed on a Ryzen 9 3900X clocked at 3.8GHz with 64GB of DDR4 RAM running at 3200 MHz. Our ARM test cases were run on a  Raspberry-Pi 4 (i.e. a Cortex-A72  at 1.5GHz and 4GB of RAM). Finally, we ran our CUDA tests on an RTX 3090 with 24GB of RAM.  

We ran our tests over the CC, BCC and FCC lattices; in each case, the lattices are chosen so that they occupy approximately the same amount of memory. More explicitly, the CC lattice has dimension $128\times128\times128$, the BCC lattice $202 \times 202 \times 202$ (but with the non-BCC points discarded) and the FCC has dimension $161\times161\times161$ (again, with the non-FCC lattice sites discarded). Lattice size is chosen so that it does not fit completely in the fastest cache memory on any given device. The lattices are chosen with roughly equivalent sizes so as to ensure one does not have an advantage over the other. 

The average reconstruction time and variance are visualized as lower diagonal matrix plots, where the group size is presented along the x-axis and the pipeline depth along the y-axis; see Figure~\ref{fig:order2} as the simplest example. Experiments on  x86\_64 and ARM measure both mean reconstruction time and variance. For the visualizations of the CUDA results, we are unable to collect the variance on a per-reconstruction basis, as CUDA batches reconstructions in groups. In these cases we omit the variance. CUDA also provides a linear texture fetch, in which case we report timing results with that optimization present --- however, when using a linear texture fetch, the concept of a group size no longer applies, as fetches are forced into groups as determined in Part I~\cite{part1}. This also limits maximum pipeline depth, as many coefficient reads may be combined into one. Additionally, linear fetches are not compatible with branch predication, as part of the texture fetch relies on computing terms of the polynomial withing a sub-region, thus, for cases in which branch predication is used, no results are presented for the linear fetch trick.

\subsection{Box Splines}
In this experiment we look at some of the box-splines found in the scientific visualisation literature --- we omit splines that emit a Cartesian coset decomposition (i.e. those that are effectively a sum of splines on two or more shifted grids). We do this for box-splines of order 2,3 and 4. 

\begin{figure}[h]
    \centering
    \includegraphics[width=0.95\linewidth]{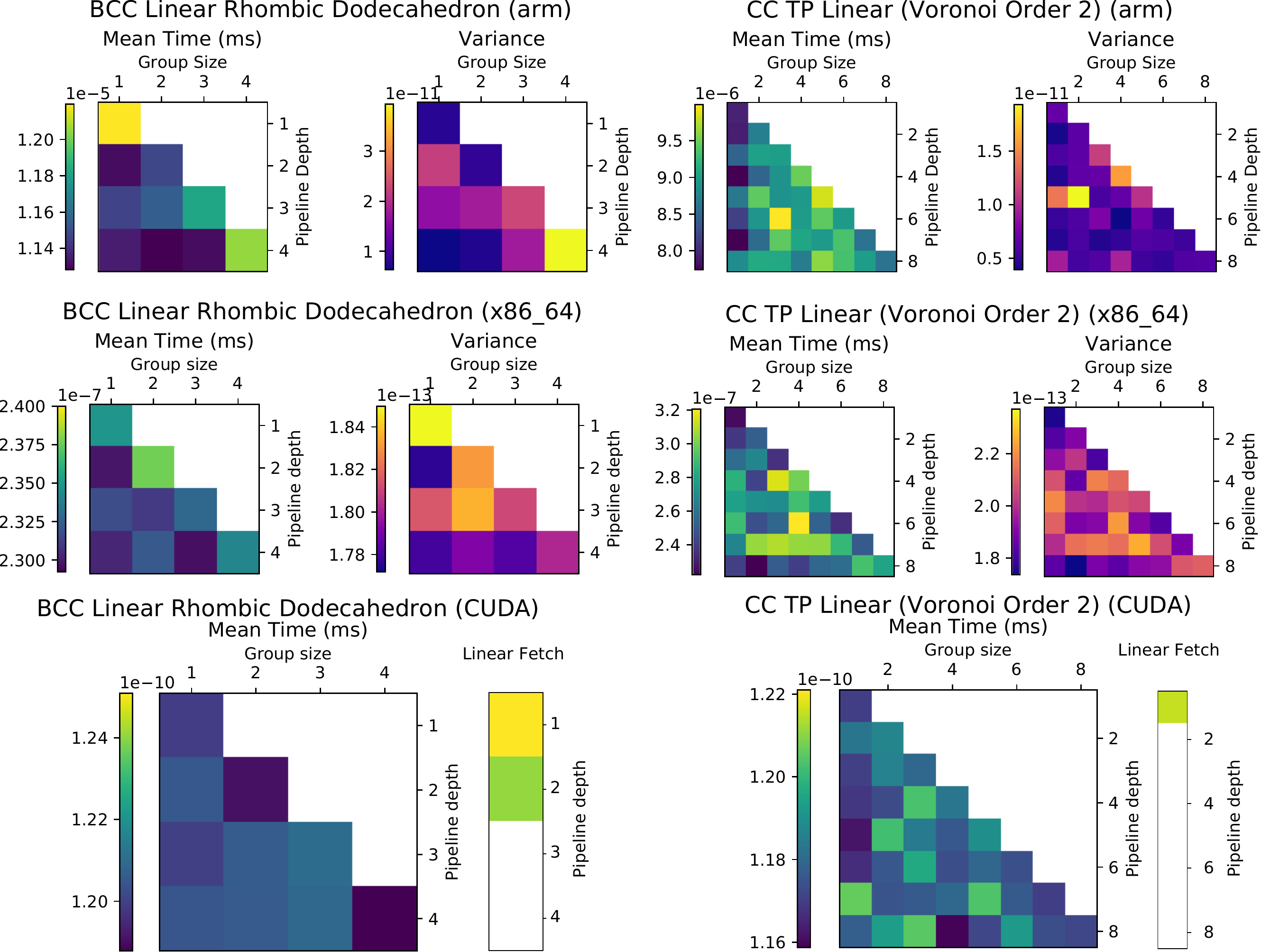} 
    \caption{Timing results for second order splines. Each square cell in the plot refers to code that has been generated with a specific value for ``pipeline depth'' and ``group size''. The mean reconstruction speed is averaged over $10^6$ trials at random locations in the volume. GPU results have no variance collected, but do include timing results for cases in which a linear texture fetch has been used to group multiple reads together. }
    \label{fig:order2}
\end{figure}
Figure~\ref{fig:order2} compares second order box-splines. There are only two cases in the literature to consider, the BCC Linear spline~\cite{entezari2004linear} and the tensor-product linear spline. Between architectures one sees orders of magnitudes of speed differences. The ARM results are the slowest, being orders of magnitude slower than
x86\_64 --- an expected result, since the Ryzen 9 is fabricated at a lower process node, includes features like branch prediction and out of order execution that are not present on the ARM chip, and is clocked higher than the Cortex CPU. The CUDA code also performed orders of magnitude faster than the x86\_64 code. Again, this is expected, since the raw parallel compute capability of the RTX 3090 is many orders of magnitude higher than that of the Ryzen 9.

The linear rhombic dodecahedon box-spline requires only four memory lookups and has a relatively simple polynomial representation compared the the tri-linear interpolant, which requires eight memory lookups and includes more terms in its polynomial representation. In our ARM tests, as expected, the linear rhombic dodecahedron spline to outperforms the tri-linear. This is a trend that is seen in the ARM code, but not the x86\_64 or GPU code. This discrepancy is likely due to the fact that there is slightly worse data locality on the BCC lattice compared to the CC --- splitting lattice cosets into separate textures has the side effect of reducing data locality. For both test splines, one sees no benefit in making use of a linear fetch. This is strange for the tensor product linear spline on the CC lattice, as the linear texture fetch is natively implemented in hardware on the card. However, our framework is not intelligent enough to simply replace this case with a single texture fetch instruction; there is a small amount of overhead in the preamble which likely leads to this case under-performing.

\begin{figure}[t]
    \centering
    \includegraphics[width=0.95\linewidth]{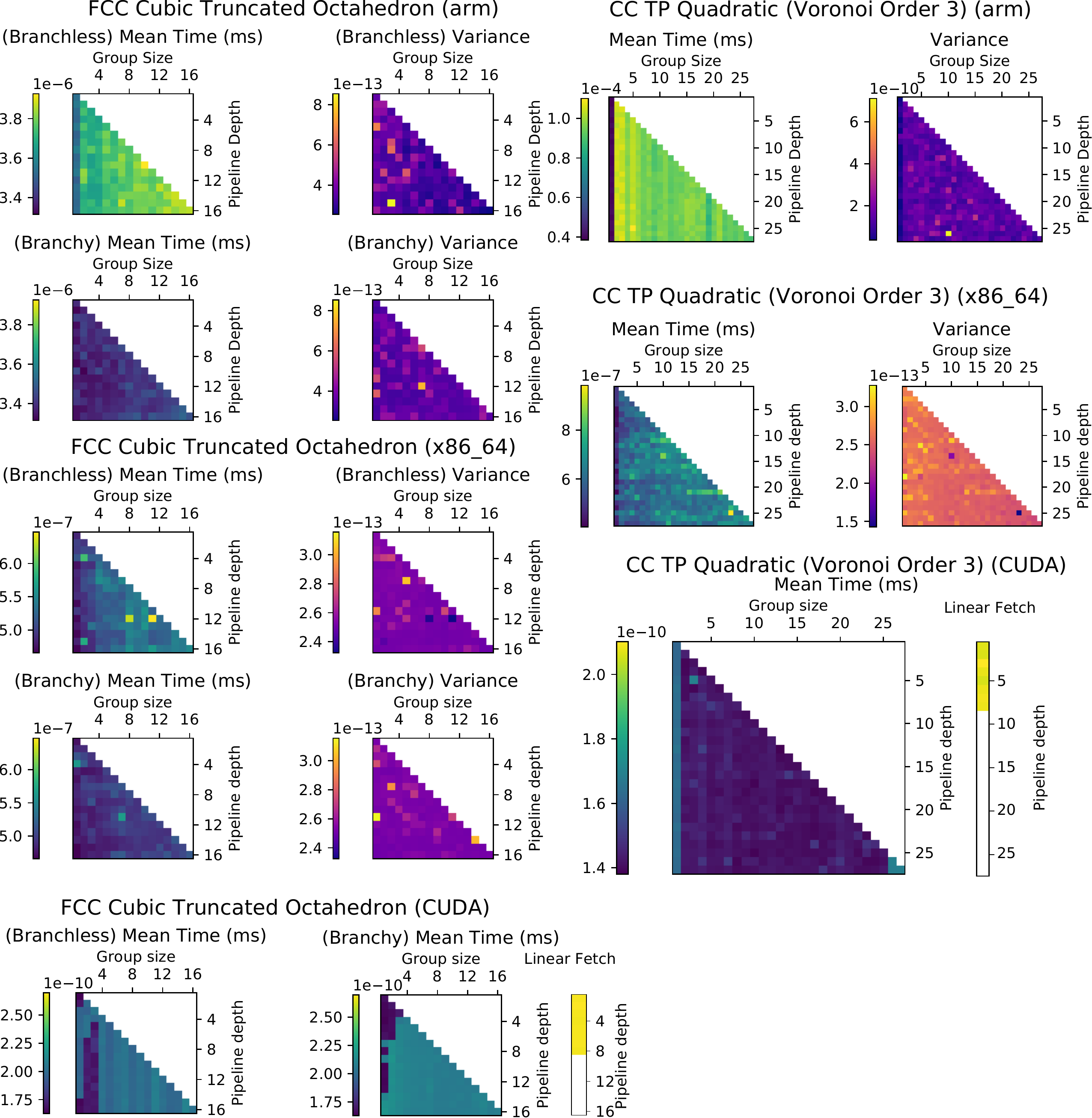} 
    \caption{Timing results for third order splines. The FCC Cubic Truncated Octahedron spline~\cite{kim2008box} emits a decomposition that requires either branch predication or branching, labelled as ``branchless'' and ``branchy'' respectively.}
    \label{fig:order3}
\end{figure}

 Figure~\ref{fig:order3} compares third order splines. This marks the first experiment in this work in which we see the use of branch predication --- the FCC cubic truncated octahedron box-spline has two representative sub-regions~\cite{kim2013efficient}, which necesscitates either a branch or branch predication. In this case, for both CPU results we see the FCC cubic spline readily outperforming the tensor product quadratic on the CC lattice. This is likely due to the simple fact that the FCC cubic spline has many fewer points in its support. Requiring fewer memory accesses on a CPU architecture where arithmetic is cheap would lead to all-around better performance.
 
 The use of branching provides insight into the effect of support size on reconstruction speed. When branching is disabled, reconstruction with the FCC cubic spline requires approximately twice the amount of computation, but this does not amount to doubling of the reconstruction time when compared to the case in which branching is enabled. This suggests that memory bandwidth has a bigger effect on reconstruction time than raw compute requirements do. This observation is present in both CPU results, but is exaggerated on x86\_64 which is likely due to its superior arithmetic performance in general. It is difficult to make a similar observation from the GPU results in this case, we will touch more on the effect branching has on the GPU when we look at the Voronoi splines, but in this specific case branching provides a tangible benefit in reconstruction speed, but only for certain parameter combinations, on average it seems to hurt performance.
 
 The tensor product quadratic spline has some interesting discrepancies. Between different CPU architectures, there is a vast difference in the effect of the group size and pipeline depth. Specifically for the ARM CPU, after a group size of 1, increasing the group size leads to better performance, but it is the opposite on the x86 CPU. However, both attain their best performance when group size is 1.
 
 Between the two splines, we see the spline with smaller support (the FCC cubic) outperforming the tensor product spline on both CPU architectures. However, we see the FCC cubic spline under-performing on the GPU, again, likely due to memory locality. However, the performance of the best case FCC cubic code is similar to the best case tensor product quadratic. In both test splines, introducing a linear fetch hurts performance.
 
\begin{figure}[t!]
    \centering
    \includegraphics[width=0.85\linewidth]{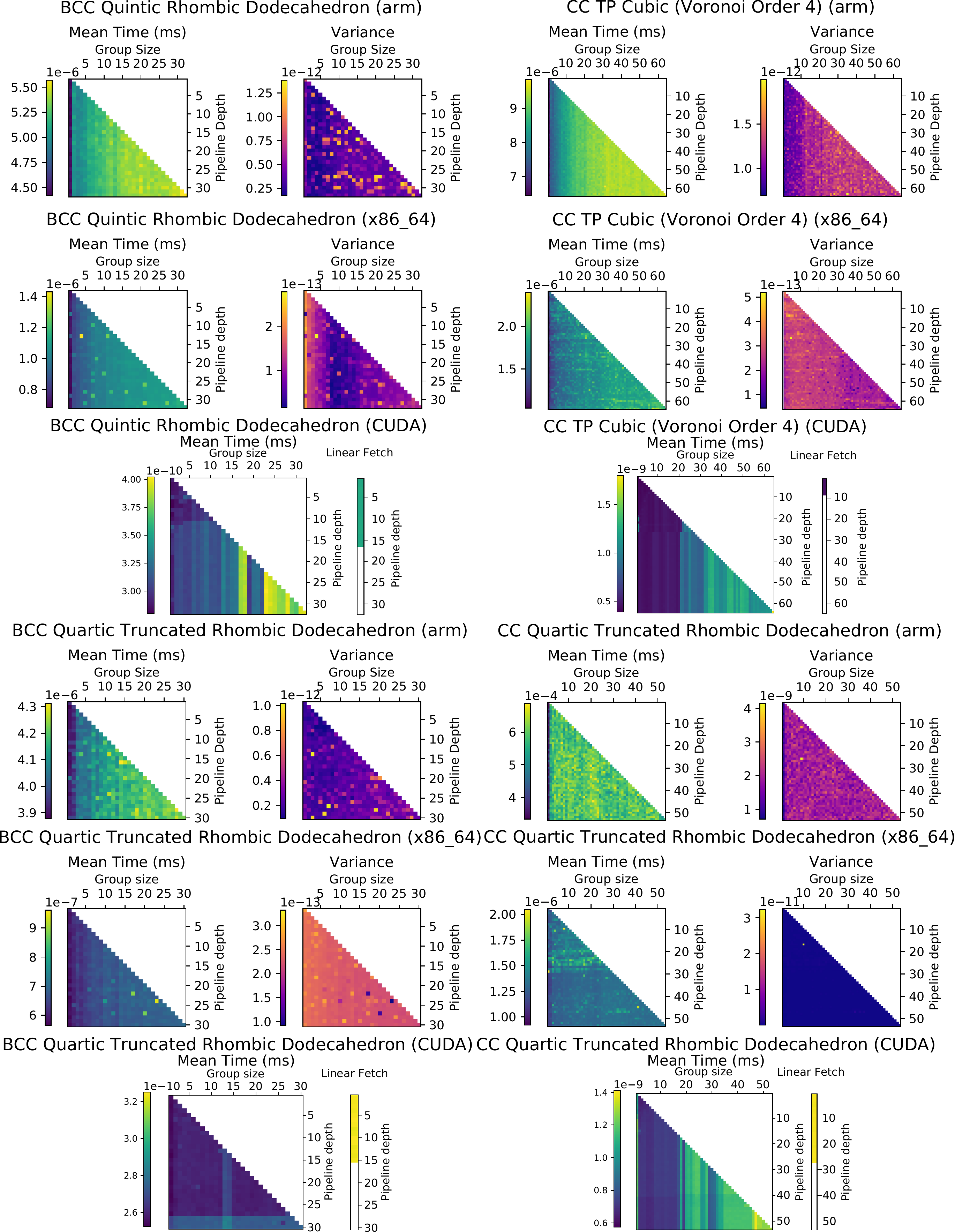} 
    \caption{Timing results for fourth order splines.}
    \label{fig:order4}
\end{figure}
 Figure~\ref{fig:order4} compares fourth order splines.~\cite{entezari2006extensions}. Most cases echo observations seen for the third order splines. The notable exceptions are certain GPU results. The tensor product cubic spline on the CC lattice marks the first instance in which the linear fetch trick improves performance. This is unsurprising, as we reduce 64 fetches to 8 via the linear fetch trick in this instance. This experiment also marks the first time a non-Cartesian spline outperforms equivalent order Cartesian splines --- both splines on the BCC lattice outperform the fourth order splines on the CC lattice, with the BCC quartic spline being fastest in all cases.

\subsection{Voronoi Splines}
The Voronoi splines are important test cases as they provide the same support size and order on all 3D lattices. As such, we consider them the most fair comparison to the tensor product splines on the Cartesian lattice (which are exactly the Voronoi splines for the Cartesian lattice). 

\begin{figure}
    \centering
    \includegraphics[width=0.95\linewidth]{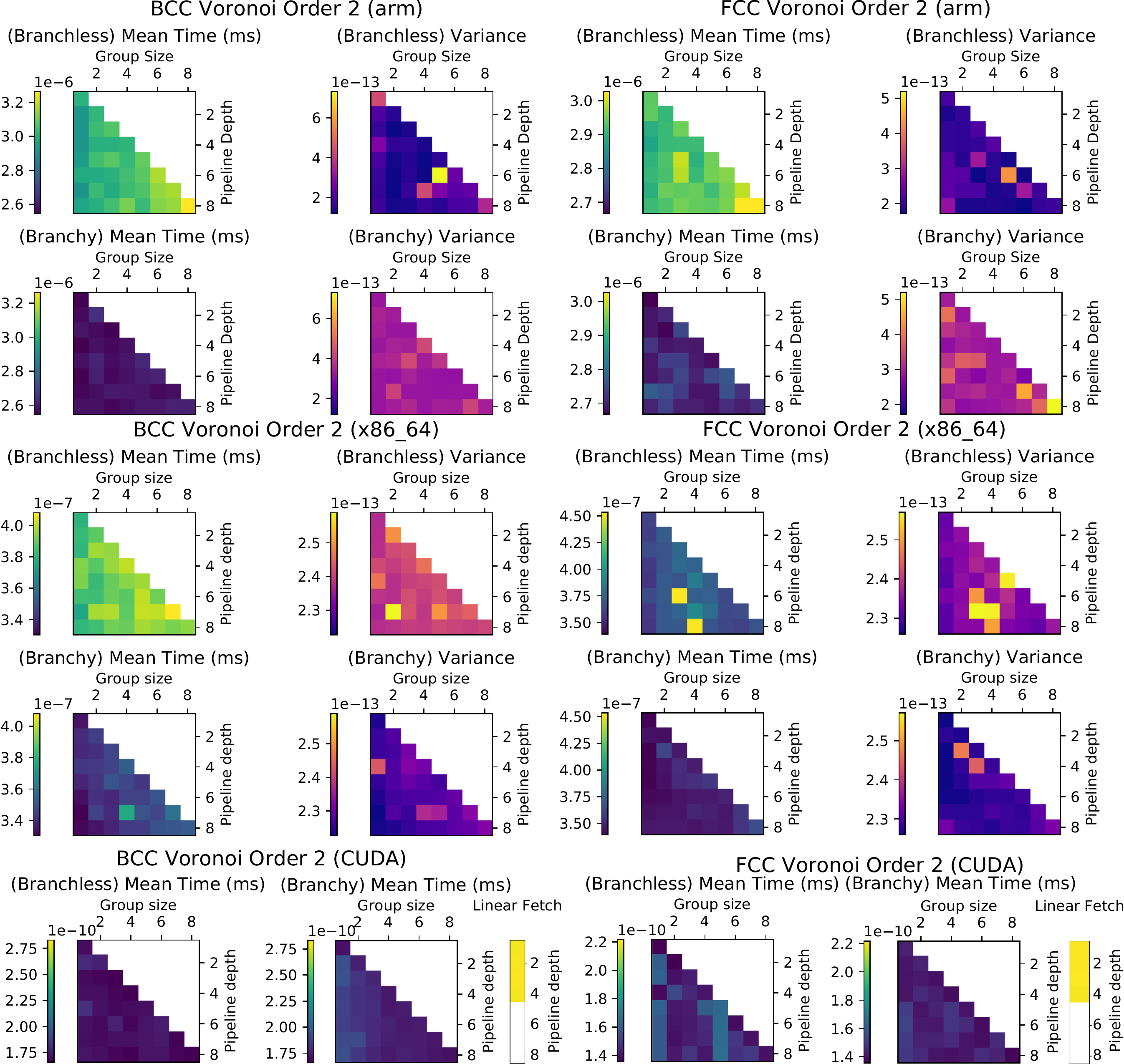}
    \caption{Timing results for the second order Voronoi splines. The results on the left were generated on an x86 machine, whereas the results on the right were generated on an ARM machine. }
    \label{fig:voronoi2}
\end{figure}

Figure~\ref{fig:voronoi2} shows the performance results for the second order Voronoi splines. In general, on the both the CPU and GPU, the non-Cartesian Voronoi splines perform on par with the tensor product linear spline (see Figure~\ref{fig:order2} as a comparison) when branchy code is generated --- they perform on the same order of magnitude with a small speed penalty likely due to the additional non-separability of their polynomial regions and memory incoherence. The effect of branching on the CPU code in this case is consistent with the observations for the FCC cubic box spline. This is many orders of magnitude faster than what is reported in the original Voronoi spline paper, in which it is cited that their renders took many days to complete~\cite{mirzargar2010voronoi, mirzargar2011quasi}. On the GPU, the effect of branching is inconclusive.  The second order BCC spline is hurt by allowing branches (on average), however the second order FCC spline is aided by the addition of branches (on average). Enabling linear fetches reduces performance.

\begin{figure}
    \centering
    \includegraphics[width=0.95\linewidth]{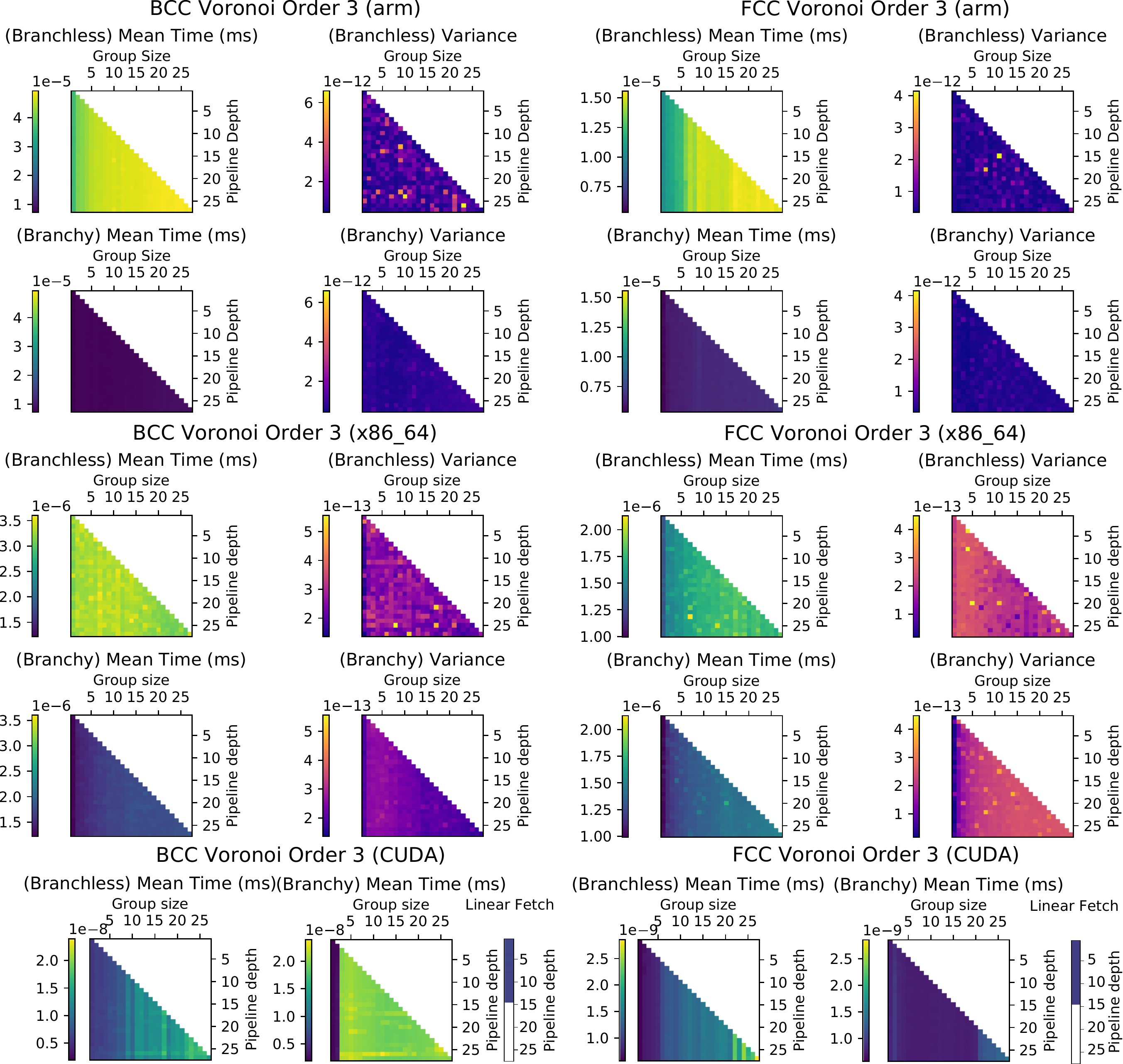}
    \caption{Timing results for the third order Voronoi splines. }
    \label{fig:voronoi3}
\end{figure}

\begin{figure}
    \centering
    \includegraphics[width=0.95\linewidth]{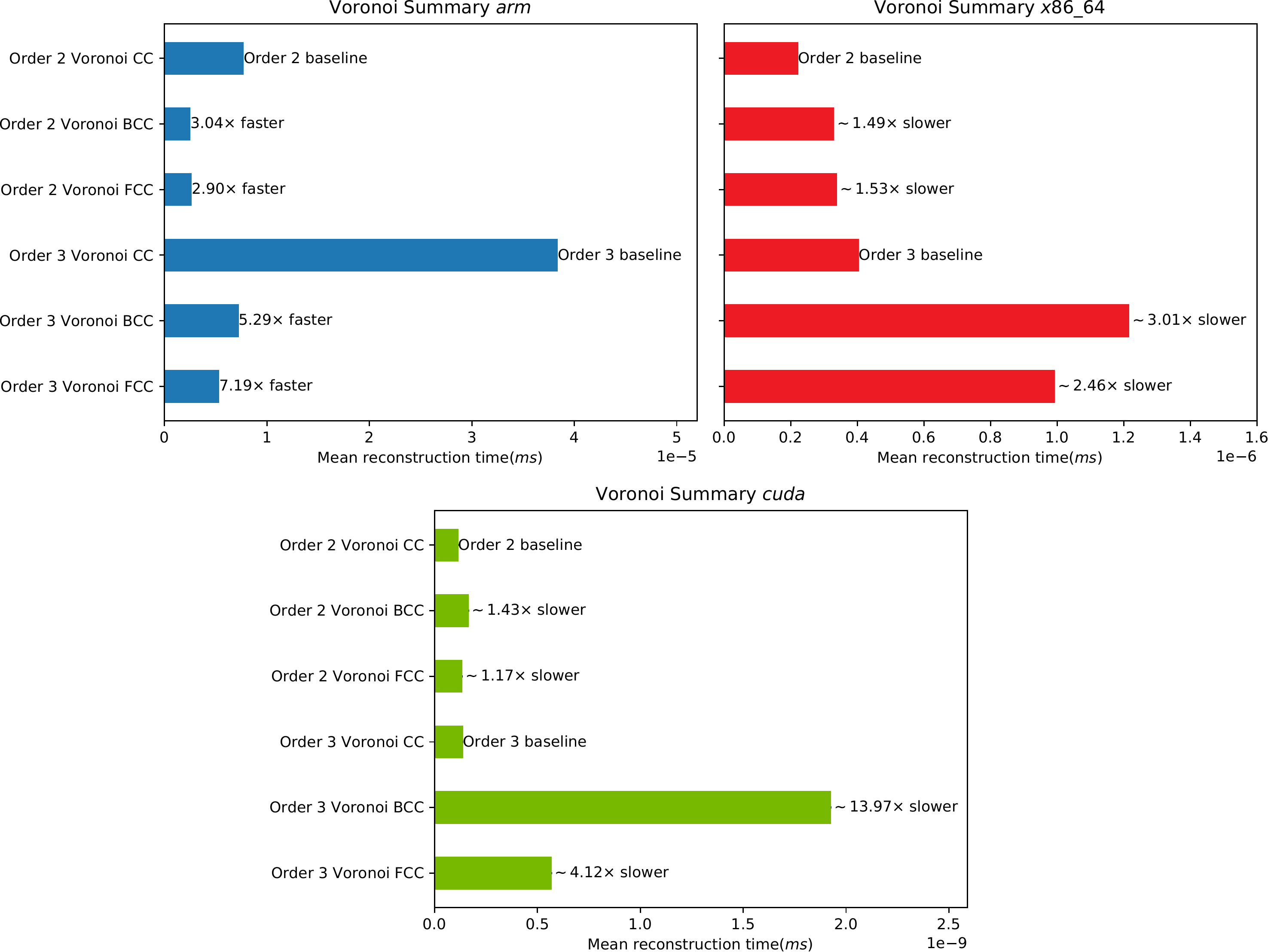}
    \caption{Best performance for Voronoi, each bar shows the relative performance to the "baseline" Cartesian Voronoi spline. }
    \label{fig:vorsum}
\end{figure}

Figure~\ref{fig:voronoi3} shows the performance for the third order Voronoi splines. On the CPU, the best case branchy code is roughly on par with the third order tensor product spline (see Figure~\ref{fig:order3}). The performance is significantly worse than the branchy case. The likely reason for this is the 7 unique sub-region polynomials the must be predicated on each evaluation. This leads to an excess of additional work that must be performed on each evaluation. On the GPU, branching increases performance on the FCC lattice on average, and reduces performance on the BCC lattice (on average). However, the best GPU performance on the BCC lattice is attained when branching is enabled. Additionally, the best performanceon the BCC lattice is on the same order of magnitude as the best performance on the FCC lattice (albeit a small factor slower). Both results are an order of magnitude slower than the GPU code for the third order tensor product spline on the Cartesian lattice (Figure~\ref{fig:order3}). Figure~\ref{fig:vorsum} shows a summary of the {\em best} performance of the different Voronoi splines (on their native lattices) on different architectures. Notably, the ARM experiments shows better relative performance on the BCC and FCC lattices compared to the x86\_64 implementations. This is likely because the ratio of compute to memory bandwidth is higher on the ARM chip --- that is, the ARM implementation has more time to "compute" while waiting for memory fetches to return. Compared to the x86\_64 implementation in which the compute to memory bandwidth ratio is much lower. This would also imply that the Quadratic polynomial that is being evaluated is somehow more complicated than those for the Voronoi spline. This would imply that more work must be done to compute the Quadratic tensor product spline than the other Voronoi splines, which is counter intuitive, since tensor product splines are notably more simple than their non-separable counterparts. Keep in mind that we expand the tensor product representation, then use a greedy Horner factorization for evaluation, which leads to more work than the original tensor product representation would have. The GPU has a higher compute to memory bandwidth ratio, but gets penalized heavier for more complicated memory reads. Moreover, this better compute to bandwidth ratio gets consumed by the branch predication, which must waste up to 7$\times$ the computation for evaluation on the BCC lattice. 

\subsection{Overall remarks}
Overall, the parameters we introduced into our code generation scheme have a large impact on the speed of resulting approximation scheme. In general, for CPU based interpolation methods enabling branches in the code generation leads to more efficient implementations --- this is an expected result, since the performance penalty of branching on CPU architectures is very small compared to the performance impact of wasting computation. The same cannot be uniformly said for GPU, there are indeed cases in which branch predicated code performs better than branch-heavy code. However, our results show that the dogma of avoiding branches in GPU code is not completely true, at least on modern Volta GPUs. On average, however, branching seems to hurt performance more than it helps performance.

In terms of polynomial grouping, larger groups tend to worse performance in most cases. This is somewhat counter-intuitive, as one would expect grouping multiple memory reads together would result in less time spent on polynomial evaluations. However, smaller grouping likely leads to better interleaving of computation and memory fetches. Keep in mind that setting the group size low may lead to lower accuracy results compared to larger group sizes, since a larger group size combines more terms during the Horner factorization, which implies lower error rates.

Pipeline depth does seem to have an effect on reconstruction speed, but it is not as pronounced as the memory fetch grouping. A good set of default parameters seems to be a low group size (simply set to 1) and a high pipeline depth (the maximum value) with branches enabled for CPU code, and disabled for GPU code. 

%% file: sections/conclusion.tex
\section{Conclusions, Limitations and Future Work}
Long has the belief been held that non-Cartesian methods are impractical because they introduce additional complexity and unacceptable computational overhead. First of all, this is not true in general, we have explicitly shown in this work that there are cases in which non-Cartesian splines out perform Cartesian splines. The matter is more nuanced than it appears, and depends on many factors including machine architecture, polynomial representation, pipelining, to name a few.

Thanks to modern advances in GPU technology and theoretical frameworks (i.e. thi    s work) we claim that the overhead for non-Cartesian interpolation is now negligible on the GPU. We should be clear, we are not stating that there is no overhead to non-Cartesian splines, there clearly is in certain cases, but we are at a point where raw computation is cheap enough to use such methods. We also show good performance on the CPU as well, but the sentiment against non-Cartesian methods has not seemed as strong for CPU based implementations in the literature. 

Additionally, by using the results from our previous work in Part I~\cite{part1}, and proposing a framework for generating efficient cross-platform implementations of spline spaces, we lift the burden of implementation away from the practitioner. While these results are likely not the absolute fastest possible, our methodology is applicable to a large number of splines, including splines that are very difficult to implement and have not received any form of an efficient implementation in the literature (i.e. the Voronoi Splines). 

Our code generation framework includes a small set of tune-able parameters that impact performance. In exploring these, we exposed the effect of these parameters on performance. Moreover, we showed that parameter choice affects performance differently depending on architecture --- this affords a practitioner more options in tuning performance on a given architecture, as it is clear that one evaluation scheme is not guaranteed to attain optimal performance over all possible compute architectures. We also provide a reference implementation of both Part I and II~\cite{fastsplinegit}. Future work is devoted to investigating alternative polynomial representations, increasing the class of interpolants to which this work is applicable (i.e. to extend beyond polynomial interpolants), incorporating gradient computations and applying additional heuristics to accelerate tensor product forms.